\documentclass[prd,amsmath,amssymb,showpacs,superscriptaddress,nofootinbib]{revtex4}

\usepackage{amssymb}
\usepackage{multirow}
\usepackage{graphicx,epsfig}
\usepackage{dcolumn}
\usepackage{bm}
\usepackage[dvipdfm,CJKbookmarks=true,unicode,colorlinks,linkcolor=blue,anchorcolor=blue,citecolor=blue,pdfborder={0 0 0}]{hyperref}
\usepackage{amsfonts}
\usepackage{keyval,graphicx}
\usepackage{textcomp,wasysym}

\begin{document}
\newcommand{\cinst}[2]{$^{\mathrm{#1}}$~#2\par}
\newcommand{\crefi}[1]{$^{\mathrm{#1}}$}
\newcommand{\crefii}[2]{$^{\mathrm{#1,#2}}$}
\newcommand{\crefiii}[3]{$^{\mathrm{#1,#2,#3}}$}
\newcommand{\HRule}{\rule{0.5\linewidth}{0.5mm}}
\newcommand{\br}[1]{\mathcal{B}#1}
\newcommand{\el}[1]{\mathcal{L}#1}
\newcommand{\ef}[1]{\mathcal{F}#1}

\parskip=3pt plus 1pt minus 1pt

\title{\boldmath Study of $\psi(3686) \rightarrow \omega K \bar{K} \pi$ decays}
\author{
\small
M.~Ablikim$^{1}$, M.~N.~Achasov$^{6}$, O.~Albayrak$^{3}$,
D.~J.~Ambrose$^{39}$, F.~F.~An$^{1}$, Q.~An$^{40}$, J.~Z.~Bai$^{1}$,
R.~Baldini Ferroli$^{17A}$, Y.~Ban$^{26}$, J.~Becker$^{2}$
% ...
\\
\vspace{0.2cm}
(BESIII Collaboration)\\
\vspace{0.2cm} {\it
$^{1}$ Institute of High Energy Physics, Beijing 100049, People's Republic of China\\
$^{2}$ Bochum Ruhr-University, D-44780 Bochum, Germany\\
$^{3}$ Carnegie Mellon University, Pittsburgh, Pennsylvania 15213, USA\\
% ...
$^{6}$ G.I. Budker Institute of Nuclear Physics SB RAS (BINP), Novosibirsk 630090, Russia\\
% ...
$^{17}$ (A)INFN Laboratori Nazionali di Frascati, I-00044, Frascati, Italy; (B)INFN and University of Perugia, I-06100, Perugia, Italy\\
% ...
$^{26}$ Peking University, Beijing 100871, People's Republic of China\\
% ...
$^{39}$ University of Rochester, Rochester, New York 14627, USA\\
$^{40}$ University of Science and Technology of China, Hefei 230026, People's Republic of China\\
% ...
}}
\date{\today}

\begin{abstract}

Based on $106$ million $\psi(3686)$ events collected with the BESIII
detector at the BEPCII, the decay $\psi(3686) \rightarrow \omega K
\bar{K} \pi$ is studied. Enhancements around $1.44$~GeV/$c^{2}$ and
the $f_{1}(1285)$ are observed in the mass spectrum of $K \bar{K}\pi$,
and the corresponding branching fractions are measured, as well as the
branching fractions of $\psi(3686) \rightarrow \omega
K^{*+}K^{-}+c.c.$ and $\psi(3686) \rightarrow \omega
\bar{K}^{*0}K^{0}+c.c.$, all for the first time.

\end{abstract}


\maketitle
\section{Introduction}

Charmonium decays play an important role in the study of the strong
interactions. The $\psi(3686)$, which was the second charmonium state
discovered~\cite{psi-discovery}, has been extensively studied both
experimentally~\cite{mark1-78,mark2-rhopi,bes2-2005,bes206g,cleo-2005}
and
theoretically~\cite{theory-expected1,theory-expected2,theory-expected3,
theory-expected4}. Perturbative QCD~\cite{exp-qcd1,exp-qcd2} predicts
that the partial widths for $J/\psi$ and $\psi(3686)$ decays into an
exclusive hadronic state $h$ are proportional to the squares of the
$c\bar{c}$ wave-function overlap at zero quark separation, which are
well determined from the leptonic widths. Since the strong coupling
constant, $\alpha_{s}$, is not very different at the $J/\psi$ and
$\psi(3686)$ masses, it is expected that the $J/\psi$ and $\psi(3686)$
branching fractions of any exclusive hadronic state $h$ are related by
\begin{eqnarray}
Q_{h}=\frac{\mathcal{B}(\psi(3686)\rightarrow h)}{\mathcal{B}(J/\psi
\rightarrow h)} \cong \frac{\mathcal{B}(\psi(3686)\rightarrow
e^{+}e^{-})} {\mathcal{B}(J/\psi \rightarrow e^{+}e^{-})} \cong
12\%.\nonumber
\end{eqnarray}
This relation defines the "$12\%$ rule", which works reasonably well
for many specific decay modes. A large violation of this rule was
observed by later
experiments~\cite{mark2-rhopi,bes2-2005,cleo-2005}, particularly in
the $\rho\pi$ decay. Recent
reviews~\cite{theory-expected3,theory-expected4}
of relevant theories and experiments conclude that current
theoretical explanations are unsatisfactory. Clearly, more
experimental results are also desirable.

A pseudoscalar gluonium candidate around 1.44 GeV/$c^{2}$, the
so-called $E/\iota(1440)$, was first observed in $p\bar{p}$
annihilation in 1967~\cite{baillon67}. Studies in different
decay modes revealed the existence of two resonant structures, the
$\eta(1405)$ and the $\eta(1475)$~\cite{pdg-2012}. The $\eta(1475)$
could be the first radial excitation of $\eta^{'}(958)$.
The L3 measurements of the $K\bar{K}\pi$ and $\eta\pi^{+}\pi^{-}$
decays in photon-photon fusion suggest that the $\eta(1405)$ has a
large gluonic content~\cite{acciarri01g}. However, CLEO with a five
times larger data sample did not confirm the L3 results, but their
upper limits are still consistent with the glueball and the radial
excitation hypotheses for the $\eta(1405)$ and
$\eta(1475)$~\cite{eta1405-cle0c}.

In this paper, the first observation of enhancements at around $1.44$
GeV/$c^{2}$ and at the $f_{1}(1285)$ resonance in the mass spectrum of
$K \bar{K}\pi$ ($K^{0}_{S}K^{+}\pi^{-}+c.c.$ and $K^{+}K^{-}\pi^{0}$)
produced in $\psi(3686) \rightarrow \omega K\bar{K} \pi$ and the
measurements of the corresponding branching fractions are reported. In
addition, the branching fractions of $\psi(3686) \rightarrow \omega
K^{*+}K^{-}+c.c.$ and $\psi(3686) \rightarrow \omega
\bar{K}^{*0}K^{0}+c.c.$ are also measured for the first time. The
analysis reported here is based on $1.06 \times 10^{8} \psi(3686)$
events collected with the BESIII detector at BEPCII.

\section{BESIII Detector}

BEPCII/BESIII~\cite{bes3} is a major upgrade of the BESII experiment
at the BEPC accelerator~\cite{bes2,bes2-p2} for studies of hadron
spectroscopy, charmonium physics, and $\tau$-charm physics
\cite{bes3phys}. The design peak luminosity of the double-ring
$e^+e^-$ collider, BEPCII, is $10^{33}$ cm$^{-2}s^{-1}$ at a beam
current of 0.93~A at the $\psi(3770)$ peak.  The BESIII detector
with a geometrical acceptance of 93\% of 4$\pi$, consists of the
following main components: 1) a small-celled, helium-based main
draft chamber (MDC) with 43 layers. The average single wire
resolution is 135~$\mu m$, and the momentum resolution for 1~GeV/$c$
charged particles in a 1~T magnetic field is 0.5\%; 2) an
electromagnetic calorimeter (EMC) made of 6240 CsI (Tl) crystals
arranged in a cylindrical shape (barrel) plus two endcaps. For
1.0~GeV photons, the energy resolution is 2.5\% in the barrel and
5\% in the endcaps, and the position resolution is 6~mm in the
barrel and 9~mm in the endcaps; 3) a Time-Of-Flight system (TOF) for
particle identification composed of a barrel part made of two layers
with 88 pieces of 5~cm thick, 2.4~m long plastic scintillators in
each layer, and two endcaps with 96 fan-shaped, 5~cm thickness, plastic
scintillators in each endcap.  The time resolution is 80~ps in the
barrel, and 110~ps in the endcaps, corresponding to a K/$\pi$ separation better than a 2
$\sigma$ for momenta below about 1~GeV/$c$; 4) a
muon chamber system (MUC) made of 1000~m$^2$ of Resistive Plate
Chambers (RPC) arranged in 9 layers in the barrel and 8 layers in
the endcaps and incorporated in the return iron of the
superconducting magnet. The position resolution is about 2~cm.

The GEANT4-based simulation software BOOST~\cite{sim-boost} includes
the geometric and material description of the BESIII detectors, the
detector response and digitization models, as well as the tracking
of detector running conditions and performance. $1.06 \times
10^{8}$ inclusive Monte Carlo (MC) events are used in our background
studies. The production of the $\psi(3686)$ resonance is simulated
with the event generator KKMC~\cite{sim-kkmc,sim-kkmc2}, while the
decays are generated with EvtGen~\cite{sim-evtgen} with known
branching fractions~\cite{pdg-2012}, and by
Lundcharm~\cite{sim-lundcharm} for unmeasured decays. The analysis
is performed in the framework of the BESIII Offline Software System
(BOSS)~\cite{ana-boss} which takes care of the detector calibration,
event reconstruction, and data storage.

\section{Event selection}

In this analysis, the $\omega$ meson is reconstructed in its
dominant decay $\omega \rightarrow \pi^{+}\pi^{-}\pi^{0}$;
$K^{0}_{S}$ and $\pi^{0}$ are reconstructed from the decays
$K^{0}_{S} \rightarrow \pi^{+}\pi^{-}$ and $\pi^0 \rightarrow \gamma
\gamma$. The final states of $\psi(3686) \rightarrow \omega
K^{0}_{S}K^{+} \pi^{-}$
and $\omega K^{+}K^{-}\pi^{0}$ are $2(\pi^{+}\pi^{-})
K^{+}\pi^{-}\gamma\gamma$  and $\pi^{+}\pi^{-}K^{+}K^{-}\gamma
\gamma \gamma \gamma$,
respectively$^{[1]}$~\footnotetext[1]{The charge-conjugate final state
is included throughout the paper unless explicitly stated.}.

Charged tracks are reconstructed from MDC hits. Each charged track
(except those from $K^{0}_{S}$ decays) is required to originate from
within $2$ cm in the radial direction and $20$ cm along the beam
direction of the run-by-run-determined interaction point. The tracks
must be within the MDC fiducial volume, $|\cos\theta| < 0.93$, where
$\theta$ is the polar angle. The information from the TOF and $dE/dx$
is combined to form a probability Prob$(K)$ (Prob$(\pi)$ or Prob$(p)$)
under a kaon (pion or proton) hypothesis. To identify a track as a
kaon, Prob$(K)$ is required to be greater than Prob$(\pi)$ and
Prob$(p)$.

Electromagnetic showers are reconstructed from clusters of energy
deposits in the EMC. The energy deposited in the nearby TOF counters
is included to improve the reconstruction efficiency and the energy
resolution. A photon candidate is a shower in the barrel region
($|\cos\theta|<0.8$) with an energy larger than $25$ MeV, or in the
endcap region ($0.86 < |\cos\theta| < 0.92$) larger than $50$ MeV,
where $\theta$ is the polar angle of the shower. The showers close
to the gap between the barrel and the endcap are poorly measured and
are thus excluded from the analysis. Moreover, the EMC timing, with
respect to the collision, of the photon candidate must be in
coincidence with collision events, i.e., $0 \le t \le 700$ ns, to
suppress electronic noise and energy deposits unrelated to the
event.

$K^{0}_{S}$ candidates are reconstructed from secondary vertex fits
to all oppositely charged track pairs in an event. The combination
with an invariant mass closest to the nominal $K^{0}_{S}$ mass
($m_{K^{0}_{S}}$) is kept. The reconstructed $K^{0}_{S}$ is used as
input for the subsequent kinematic fit.

\section{$\psi(3686) \rightarrow \omega K^{0}_{S} K^{+} \pi^{-}+c.c.$}

Event candidates should have a $K^{0}_{S}$, four charged tracks with
zero net charge, and two or more photons. At least one charged track
is positively identified as a kaon. When more than one kaon is
identified, the one with the maximum Prob$(K)$ is chosen. The
$\psi(3686) \rightarrow \pi^{+}\pi^{-}K^{0}_{S}K^{+}\pi^{-}\gamma
\gamma$ candidates are subjected to a four-constraint (4C) kinematic
fit provided by four-momentum conservation to reduce background and
to improve the mass resolution. For events with more than two
photons, the combination with the minimum $\chi^{2}$ value of the
fit (denoted as $\chi^{2}_{4C}$) is retained, and $\chi^{2}_{4C}$ is
required to be less than $40$. For $\pi^{0}$ candidates formed from
a photon pair, the invariant mass of the photon pair must be within
the range $|M_{\gamma\gamma}-m_{\pi^{0}}|<0.02$ GeV$/c^{2}$. The
combination of $\pi^{+} \pi^{-} \pi^0$ with an invariant mass
closest to the nominal $\omega$ mass ($m_{\omega}$) is chosen as an
$\omega$ candidate. Figure~\ref{fig:w-x1440-kskp}(a) shows the
distribution of the $M_{\pi^{+}\pi^{-}}$ versus the
$M_{\pi^{+}\pi^{-}\pi^0}$ invariant mass, where a $\omega-K^{0}_{S}$
cluster corresponding to $\psi(3686)\rightarrow\omega
K^{0}_{S}K^{+}\pi^{-}$ is apparent. Events are kept for further
analysis if the $\pi^+ \pi^-$ invariant mass is in the range $0.489$
GeV$/c^{2} <M_{\pi^{+} \pi^{-}}<0.505$ GeV$/c^{2}$ and the $\pi^+
\pi^- \pi^0$ mass in the range $0.743$ GeV$/c^{2}<M_{\pi^{+}
\pi^{-}\pi^0}<0.823$ GeV$/c^{2}$. Events in the $K^{0}_{S}$ sideband
range ($0.012$ GeV$/c^{2} <|M_{\pi^{+}\pi^{-}}- m_{K^{0}_{S}}|<0.02$
GeV/$c^{2}$) or the $\omega$ sideband ($0.06$ GeV$/c^{2}
<|M_{\pi^{+}\pi^{-}\pi^{0}}-m_{\omega}|<0.10$ GeV$/c^{2}$) are used
to estimate the background. In addition, to veto the
$\psi(3686)\rightarrow \pi^{+}\pi^{-} J/\psi $ background events,
the recoil mass against $\pi^+\pi^-$ is required to satisfy
$|M^{recoil}_{\pi^{+}\pi^{-}}-m_{J/\psi}|>0.007$ GeV$/c^{2}$.

\subsection{Branching fractions for $\psi(3686) \to \omega X(1440)\rightarrow \omega
  K^{0}_{S}K^{+}\pi^{-}+c.c.$}\label{sect-x1440}

\begin{figure}[htbp]
  \centering
\includegraphics[width=0.45\textwidth]{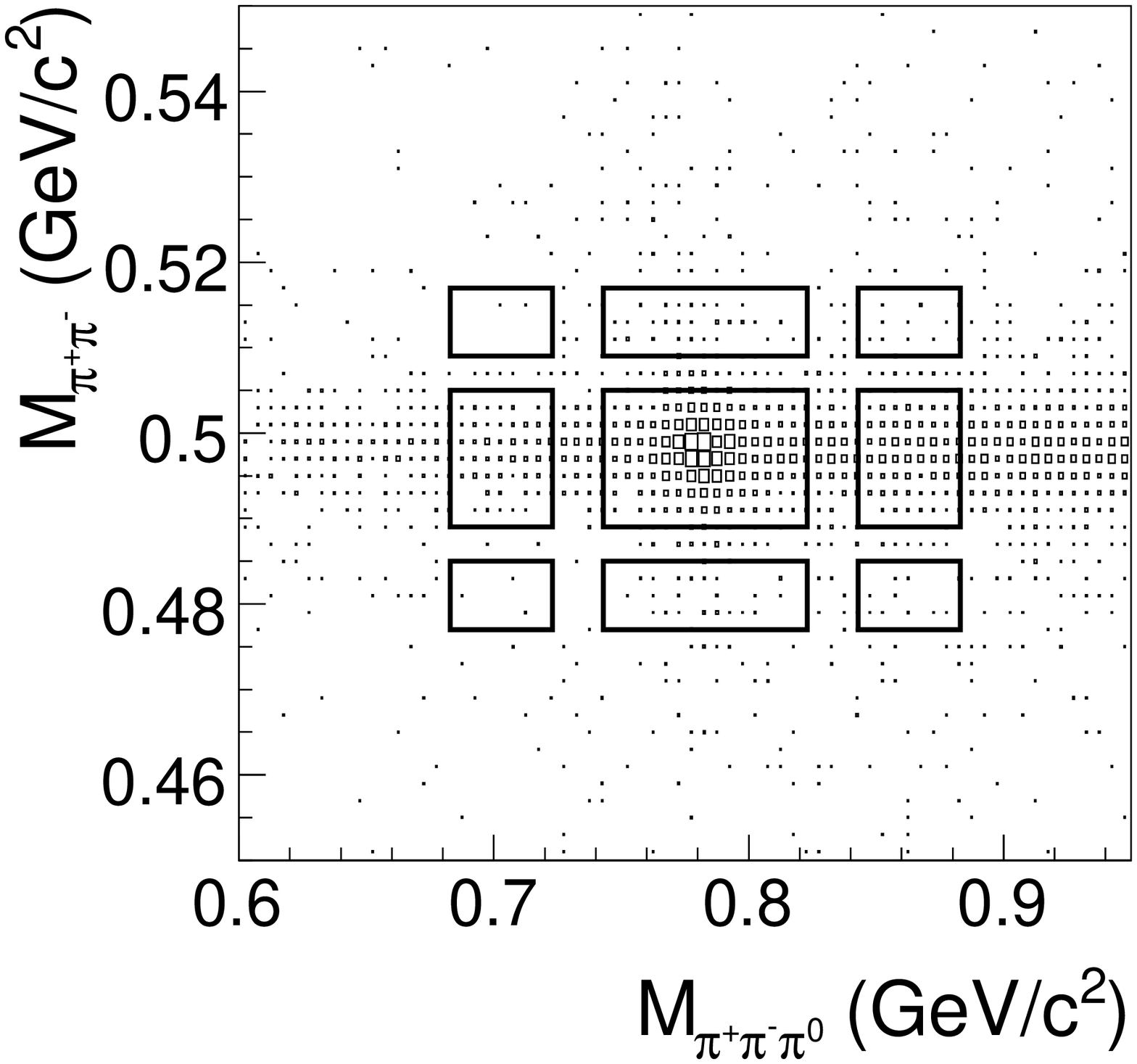}
\includegraphics[width=0.45\textwidth]{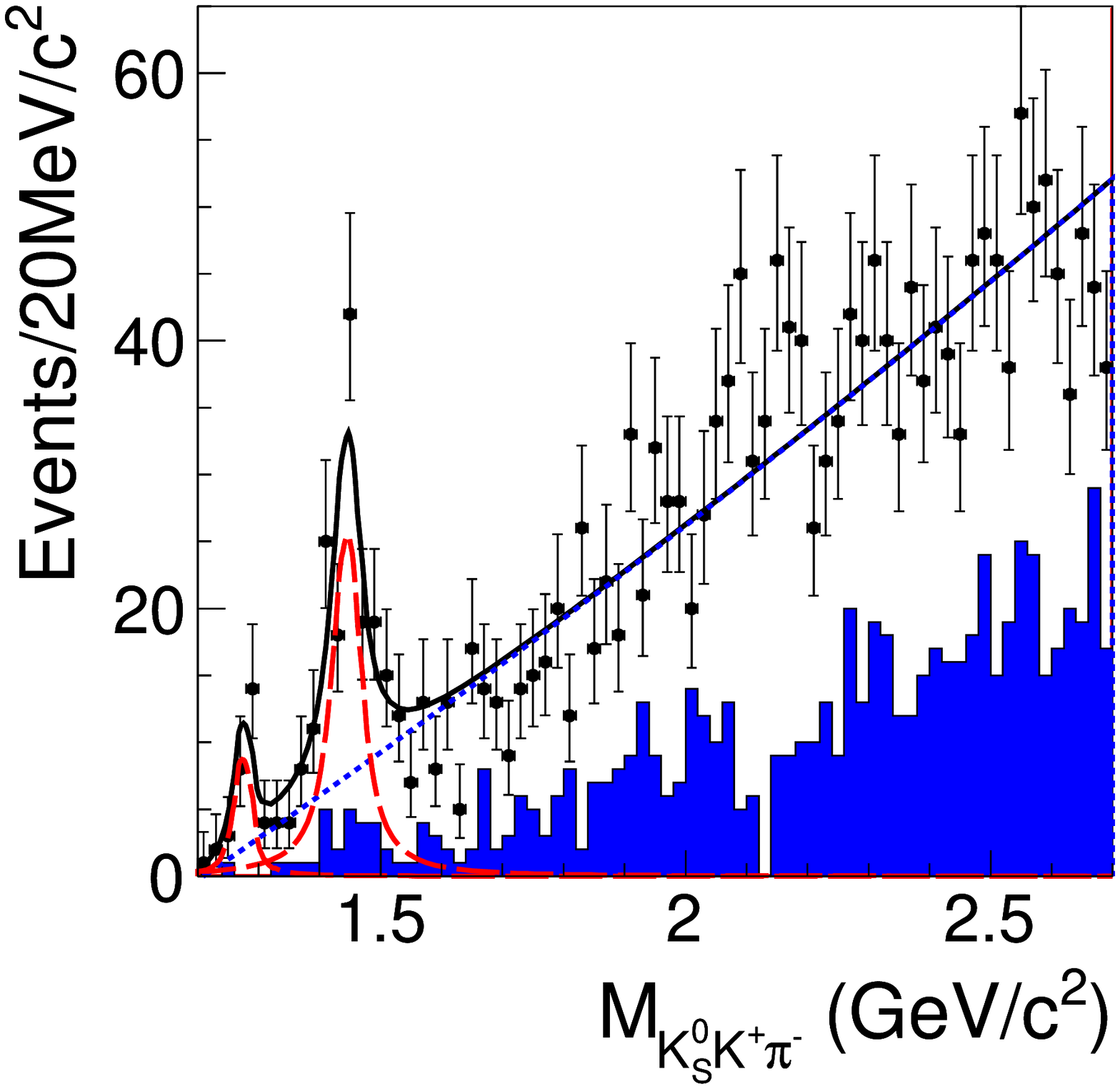}
\put(-270,170){\bf~(a)} \put(-150,170){\bf~(b)} \caption{(a)
Distribution of $M_{\pi^{+}\pi^{-}}$ versus
  $M_{\pi^{+}\pi^{-}\pi^{0}}$, where the boxes represent the
  $K^{0}_{S}$ and $\omega$ signal region and sideband regions.
  (b) The $K^{0}_{S} K^{+}\pi^{-}$ invariant-mass distribution for
  $\psi(3686) \rightarrow \omega K^{0}_{S}K^{+}\pi^{-}$
  candidate events. The shaded histogram is the background contribution
  estimated from the $K^{0}_{S}$ and $\omega$ sidebands minus that of the corner ranges.
} \label{fig:w-x1440-kskp}
\end{figure}

Figure~\ref{fig:w-x1440-kskp}(b) shows the invariant mass
$M_{K^{0}_{S}K^{+}\pi^{-}}$ for selected events, where a peak around
1.44 GeV/$c^{2}$ (denoted as X(1440), since we do not distinguish
$\eta(1405)$ and $\eta(1475)$ from this analysis) and the $f_1(1285)$
are evident. To verify that the observed peaks originate from
the process $\psi(3686) \rightarrow \omega K^{0}_{S}K^{+}\pi^{-}$, the
backgrounds are investigated from both data sideband and inclusive MC
events. The non-$K^{0}_{S}$ and non-$\omega$ backgrounds, estimated by
using events in the $K^{0}_{S}$ and $\omega$ sideband regions, and
normalized according to the ratio of MC events falling into the
sidebands to the signal region, are shown as the shaded histogram in
Fig.~\ref{fig:w-x1440-kskp}(b). No evident peak around $1.44$
GeV/$c^{2}$ is seen. Other potential $\psi(3686)$ decay backgrounds
are checked with $106$ million $\psi(3686)$ inclusive MC events. The
main backgrounds come from the decays of $\psi(3686) \rightarrow
\rho^{\pm}K^{*\mp}K^{*0}+\rho^{0}K^{*\pm}K^{*\mp} \rightarrow
\pi^{\pm} \pi^{\mp} \pi^{0} K^{0}_{S} K^{\pm} \pi^{\mp}$, which
don't form a peak
in the $M_{K^{0}_{S}K^{+}\pi^{-}}$ spectrum. The background
from the $e^+e^- \to q\bar{q}$ continuum process is studied by using
data collected at the center-of-mass energy of $3.65$ GeV. Continuum
backgrounds are found to be small and uniformly distributed in the mass
spectrum.

An unbinned maximum-likelihood fit to $M_{K^{0}_{S}K^{+}\pi^{-}}$ is
applied to determine the number of $\omega X(1440)$ and $\omega
f_1(1285)$ events. The fit includes three components: $X(1440)$,
$f_1(1285)$, and background.  Both $X(1440)$, and $f_1(1285)$ are
represented by Breit-Wigner (BW) functions convoluted with a
Novosibirsk mass-resolution function and multiplied by an efficiency
curve. Both the mass resolution and the efficiency curve are
determined from MC simulations. The background shape is described by a
2nd-order Chebychev polynomial function.  The mass and width of
$f_{1}(1285)$ are fixed at the known values
~\cite{pdg-2012,omegaetapipi-bes3}, while those for the $X(1440)$ are
allowed to float in the fit. The solid curve in
Fig.~\ref{fig:w-x1440-kskp}(b) shows the fit results. The
goodness-of-fit is $\chi^{2}/ndf=81.0/75=1.08$, which indicates a
reasonable fit. The fit yields $109.2\pm 18.0$ events for the
$X(1440)$ with a statistical significance of $9.5 \sigma$, and
$21.6 \pm 7.0$ events for the $f_{1}(1285)$ with the
statistical significance of $4.5 \sigma$. The significance is
determined by the change of the log-likelihood value and the degrees of
freedom in the fit with and without assuming the presence of a signal.
The mass and width of the $X(1440)$ are determined to be $M=1452.2\pm
5.2$ MeV$/c^{2}$ and $\Gamma=51.6 \pm 12.1$ MeV/$c^2$. The contribution of
background estimated from the sideband is $13.4\pm 6.1$ for the
$X(1440)$ and $0.1\pm 2.4$ for the $f_{1}(1285)$.  In
the analysis, these backgrounds are ignored.

With the MC-determined efficiencies (where the final-state particles
distribute uniformly over the phase space) of $(10.41\pm0.14)\%$ for
$\psi(3686)\rightarrow \omega X(1440)$ and $(10.88\pm0.15)\%$ for
$\psi(3686) \rightarrow \omega f_{1}(1285)$, the product of branching
fractions are calculated to be
\begin{eqnarray}
\mathcal{B}(\psi(3686) \rightarrow \omega X(1440))\cdot \mathcal{B}(
X(1440) \rightarrow K^{0}_{S} K^{+} \pi^{-}) & = &(1.60 \pm
0.27~({\rm stat.})) \times 10^{-5}, \nonumber
\end{eqnarray}
\begin{eqnarray}
\mathcal{B}(\psi(3686) \rightarrow \omega f_{1}(1285))\cdot
\mathcal{B}( f_{1}(1285) \rightarrow K^{0}_{S} K^{+} \pi^{-}) & =
&(3.02 \pm 0.98~({\rm stat.})) \times 10^{-6}. \nonumber
\end{eqnarray}
Systematic uncertainties are discussed in
Section~\ref{systmatic-error-analysis}. Since the $f_1(1285)$ peak
is not significant, an upper limit at a $90\%$
confidence level (C.L.) on $N_{sig}$ is determined using a Bayesian
method~\cite{pdg-2012}
by finding the value $N^{UP}_{sig}$ such that
\begin{eqnarray}
\frac{\int^{N^{UP}_{sig}}_{0}LdN_{sig}}{\int^{\infty}_{0}LdN_{sig}}=0.90,\nonumber
\end{eqnarray}
where $N_{sig}$ is the number of signal events, and $L$ is the value
of the likelihood as a function of $N_{sig}$. An upper limit at the
$90\%$ C.L of 31 $f_{1}(1285)$ events is obtained.


\subsection{Branching fractions of $\psi(3686) \to \omega K^{*+}K^{-}+c.c.$
  and $\psi(3686) \to \omega \bar{K}^{*0}K^{0}$}~\label{sect-ppp}

\begin{figure}[htbp]
  \centering
\includegraphics[width=0.45\textwidth]{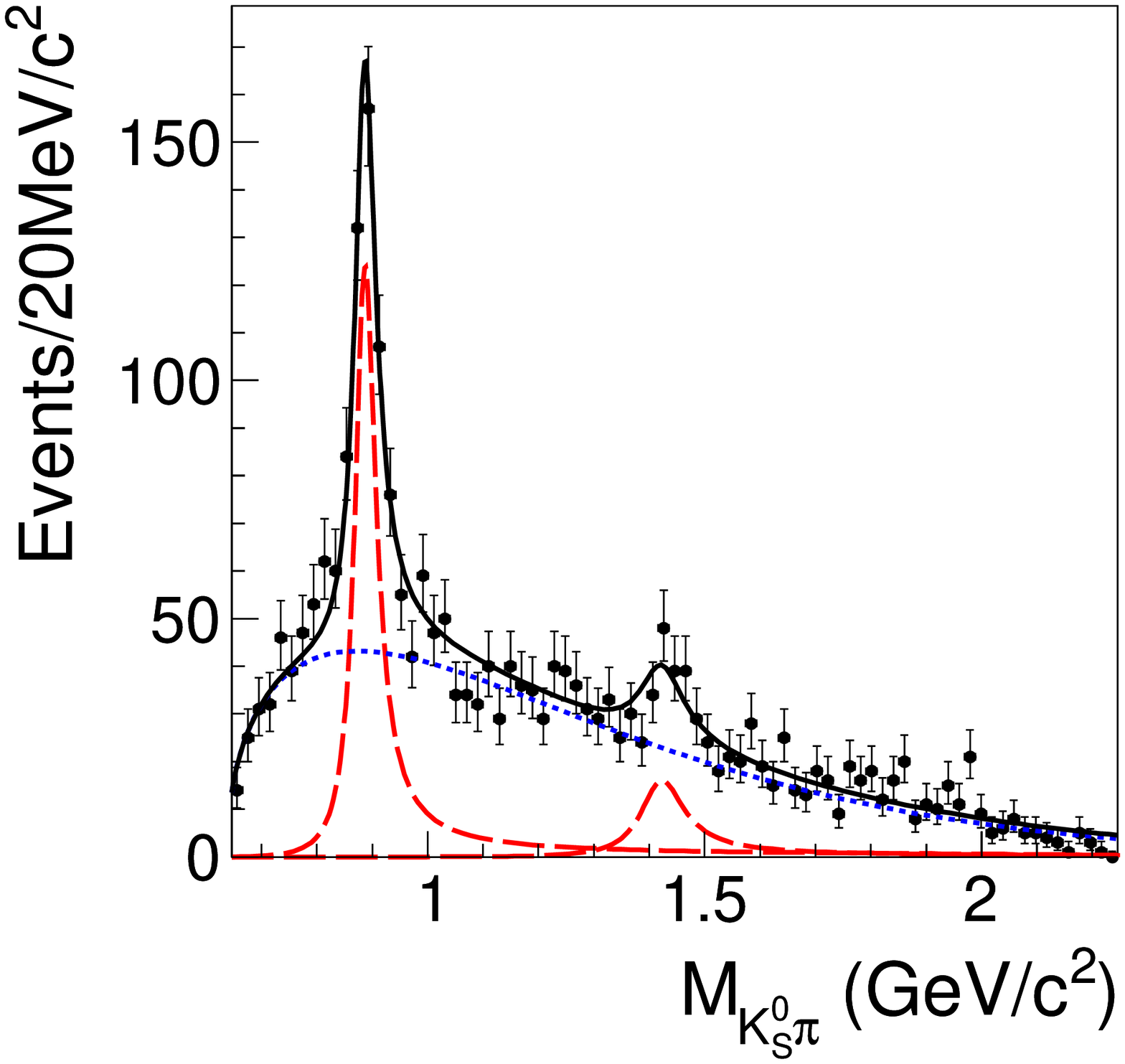}
\includegraphics[width=0.45\textwidth]{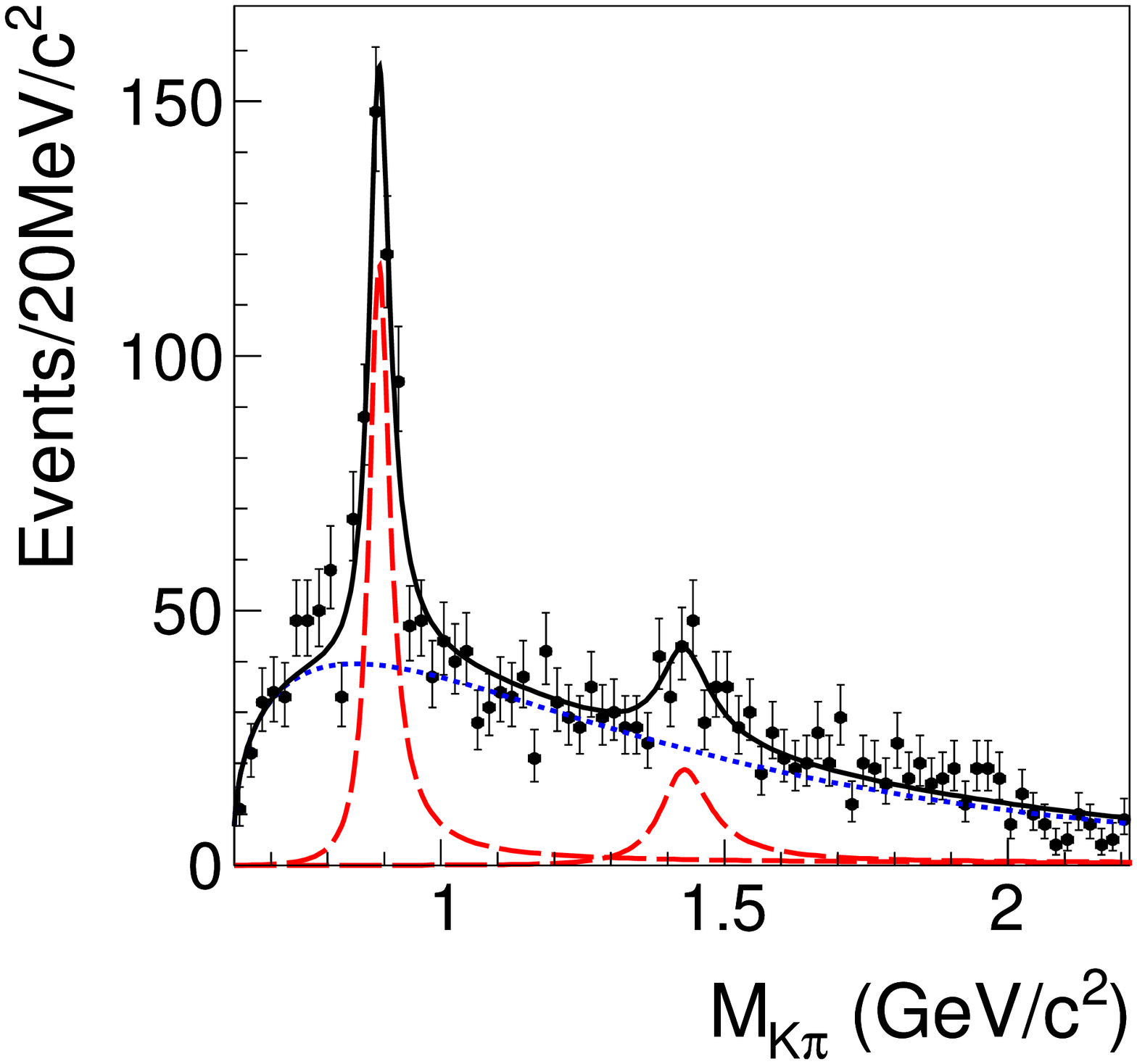}
\put(-265,165){\bf~(a)} \put(-60,165){\bf~(b)}\\
\caption{(a) The $M_{K^{0}_{S} \pi^{\pm}}$ distribution.
  (b) The $M_{K^{\mp} \pi^{\pm}}$ distribution.
  Solid curves show the fit results;
  dashed lines are for signals; and
  dotted lines are for backgrounds.}
\label{fig:fit-k892-wkskp-signal}
\end{figure}

Figure~\ref{fig:fit-k892-wkskp-signal} shows the
$M_{K^{0}_{S}\pi^{\pm}}$ and $M_{K^{\mp}\pi^{\pm}}$ distributions,
where the $K^{*}(892)$ and a peak at $1.43$ GeV/$c^2$ are
evident in both distributions. The solid lines in
Fig.~\ref{fig:fit-k892-wkskp-signal} show the results of an unbinned
maximum likelihood fit with three components: $K^{*}(892)$,
$K^{*}_{2}(1430)$, and background. Both the $K^{*}(892)$ and
$K^{*}_{2}(1430)$ are described by acceptance-corrected BW
functions. The BW
function~\cite{bwk892_cleo,bwk892_belle,bwk892} is
\begin{eqnarray}
  F_{BW}(s)=\frac{M\Gamma(s)}{(s^2-M^2)^2+M^2\Gamma(s)^2},
  \label{equ:bw-for-k892}
\end{eqnarray}
where $\Gamma(s)=\Gamma (\frac{M}{s})^{2}(\frac{q}{q_{0}})^{2L+1}$,
$M$ and $\Gamma$ are the $K^{*}$ mass and width, $q$ is the
$K^{0}_{S}$ ($K^{\mp}$) momentum in the $K^{*}$ rest-frame, $q_{0}$
is the $q$ value at $s=M$, and $L$ is the relative orbital angular
momentum of $K^{0}_{S} \pi^{\pm}$ ($K^{\mp}\pi^{\pm}$). Here the
$K^{*}(892)$ and $K^{*}_{2}(1430)$ peaks are described with $P$-wave
($L=1$) and $D$-wave ($L=2$) BW functions, respectively. The mass
and width of the $K^{*}(892)$ are floating, and those of
$K^{*}_{2}(1430)$ are fixed to the world average values
~\cite{pdg-2012}.
The backgrounds are described by the
function~\cite{bwk893_prd1984,bg-plb436}
\begin{eqnarray}
  F_{BG}(s)=(s-m_{t})^{c} e^{-d s-e s^{2}},
  \label{equ:background-for-k892}
\end{eqnarray}
where $m_{t}$ is the threshold mass for $K^{0}_{S} \pi^{\pm}$
($K^{\mp}\pi^{\pm}$) and $c$, $d$ and $e$ are free parameters.

The fit to the $M_{K^{0}_{S} \pi^{\pm}}$ distribution yields $502.0\pm
56.4$ $K^{*\pm}(892)$ events, and $128.5\pm30.0$ $K^{*\pm}_{2}(1430)$
events with a statistical significance of $4.4\sigma$. The mass and
width of the $K^{*\pm}(892)$ determined in the fit are $888.0\pm 2.5$
MeV$/c^{2}$ and $48.0\pm 6.5$ MeV/$c^2$, respectively. The fit to
$M_{K^{\mp}\pi^{\pm}}$ yields $446.2 \pm 47.4$ $K^{*0}(892)$ events,
and $164.2 \pm 34.2$ $K^{*0}_{2}(1430)$ events with a statistical
significance of $4.6\sigma$. The mass and width of the $K^{*0}(892)$
are determined to be $M=893.9\pm2.2$ MeV$/c^{2}$ and $\Gamma
=45.0\pm5.7$ MeV/$c^2$. When the peak at 1430 MeV/$c^2$ is fitted to
an $S-$wave $K^{*}_{0}(1430)$ or a $P-$wave $K^{*}(1410)$, the fit
qualities degrade for both fits. The $K^{*}_{2}(1430)$ can be
distinguished from $K^{*}_{0}(1430)$ and $K^{*}(1410)$ with
log-likelihood values worse by $173.9$ and $173.7$, respectively, in
the $K^{0}_{S} \pi^{\pm}$ decay, while the log-likelihoods are worse by
$170.6$ and $169.7$, respectively, in the $K^{\mp}\pi^{\pm}$ decay.

The peaking backgrounds for $K^{*}(892)$ and $K^{*}_{2}(1430)$ are
studied using the $K^{0}_{S}$ and $\omega$ sidebands.
In $K^{0}_{S} \pi^{\pm}$, the contributions from background events
are estimated to be $105.6\pm 21.6$ for $K^{*\pm}(892)$, which are
subtracted from the total signal yield, while no evident peaking
backgrounds for $K^{*\pm}_{2}(1430)$ are seen. Combining the numbers
of signal events with the detection efficiency of $(9.58\pm0.08)\%$
and $(9.18\pm0.07)\%$ determined from MC simulation of
$\psi(3686)\rightarrow \omega K^{*+}(892)K^{-}$ and
$\psi(3686)\rightarrow \omega K^{*+}_{2}(1430)K^{-}$, respectively,
their corresponding branching fractions are determined to be
\begin{eqnarray}
 \mathcal{B}(\psi(3686) \rightarrow \omega K^{*+}(892) K^{-}) & = &(1.89\pm 0.29~({\rm stat.}))
  \times 10^{-4},\nonumber
\end{eqnarray}
\begin{eqnarray}
 \mathcal{B}(\psi(3686) \rightarrow \omega K^{*+}_{2}(1430) K^{-}) & = &(6.39\pm 1.50~({\rm stat.})) \times 10^{-5}.\nonumber
\end{eqnarray}

In the $K^{\mp}\pi^{\pm}$ mode, after subtracting $90.2 \pm 18.3$ background
events for $K^{*0}(892)$ and $48.5\pm23.2$
for $K^{*0}_{2}(1430)$, and using the MC-determined
detection efficiencies of $(9.66\pm0.08)\%$ and $(9.08\pm0.07)\%$, respectively,
the branching fractions are
\begin{eqnarray}
  \mathcal{B}(\psi(3686) \rightarrow \omega \bar{K}^{*0}(892) K^{0}) &
  = &(1.68\pm 0.25~({\rm stat.}))
  \times 10^{-4}, \nonumber
\end{eqnarray}
\begin{eqnarray}
  \mathcal{B}(\psi(3686) \rightarrow \omega \bar{K}^{*0}_{2}(1430) K^{0}) & = &(5.82\pm 2.08~({\rm stat.}))
  \times 10^{-5}.\nonumber
\end{eqnarray}

\section{$\psi(3686) \rightarrow \omega K^{+} K^{-} \pi^{0}$}

Candidate events must have four charged tracks with zero net charge,
and four or more photons. At least two charged tracks should be identified as kaons.
When more than two charged tracks are identified as kaons
the two oppositely charged tracks with the largest Prob($K$) are
chosen. To reject the two backgrounds with $3\gamma$ or $5\gamma$ in the
final state, the $\chi^{2}_{4C}$ of a four-constraint kinematic fit
in the hypothesis of $\psi(3686) \rightarrow K^{+} K^{-} \pi^{+}
\pi^{-} 4\gamma$ is required to be less than those of the $K^{+}K^{-}
\pi^{+} \pi^{-} 3 \gamma$, and $K^{+} K^{-} \pi^{+} \pi^{-} 5
\gamma$ hypotheses. For all possible combinations, a six-constraint
kinematic (6C) fit (besides the initial $\psi(3686)$ four-momentum,
two $\pi^{0}$ masses are also used as constraints) is performed, and the one
with the least $\chi^{2}_{6C}$ is chosen, and $\chi^{2}_{6C}<100$ is
required. The combination of $\pi^{+}\pi^{-}\pi^{0}$ with the
invariant mass closest to the nominal $\omega$ mass is selected as an
$\omega$ candidate; the invariant mass $M_{\pi^{+}\pi^{-}\pi^{0}}$
must be in the region $0.743$ GeV$/c^{2}<M_{\pi^{+}
\pi^{-}\pi^0}<0.823$ GeV$/c^{2}$. Background events from $\psi(3686)
\rightarrow \pi^{+}\pi^{-} J/\psi$ and $\psi(3686) \rightarrow
\pi^{0}\pi^{0} J/\psi$ decays are removed by requiring
$|M^{recoil}_{\pi^{+}\pi^{-}}-m_{J/\psi}|>0.007$ GeV/$c^{2}$ and
$|M^{ recoil}_{\pi^{0}\pi^{0}}-m_{J/\psi}|>0.06$ GeV/$c^{2}$.
Background events from $\psi(3686) \rightarrow \gamma \gamma J/\psi$
are rejected by requiring $|M^{recoil}_{\gamma
  \gamma}-m_{J/\psi}|>0.05$ GeV/$c^{2}$.

\subsection{Branching fractions for $\psi(3686) \rightarrow \omega
  X(1440) \rightarrow \omega K^{+} K^{-} \pi^{0}$}~\label{sect-x1440-kkp0}
After the above event selection, the distribution of
$M_{K^{+}K^{-}\pi^{0}}$ is shown in Fig.~\ref{fig:x1440-wkkp0}(a),
where the $X(1440)$ is clearly seen and the $f_1(1285)$ is evident.
Non-$\omega$ background, estimated from the $\omega$ sideband
($0.06$ GeV$/c^{2}<|M_{\pi^{+}\pi^{-}\pi^{0}}-m_{\omega}|<0.10$
GeV$/c^{2}$), is shown as the shaded histogram and doesn't
form a peak
in the $M_{K^{+}K^{-}\pi^{0}}$ spectrum.

The results of an unbinned maximum-likelihood fit to
$M_{K^{+}K^{-}\pi^{0}}$, which is similar to the $M_{K^{0}_{S} K \pi}$
fit described in Section~\ref{sect-x1440}, is shown as a solid line in
Fig.~\ref{fig:x1440-wkkp0}. The goodness-of-fit is
$\chi^{2}/ndf=63.6/67=0.95$, which indicates a reasonable fit. The fit
yields $81.8\pm14.7$ $X(1440)$ events with a statistical significance
of $9.3 \sigma$, and $9.5 \pm 5.3$ $f_{1}(1285)$ events with a
statistical significance of $3.2\sigma$. The mass and width of the
$X(1440)$ determined from the fit are $M=1452.7 \pm 3.8$ MeV$/c^{2}$
and $\Gamma=36.8 \pm 10.5$ MeV/$c^2$, respectively. The non-$\omega$
contributions estimated from the $\omega$ sideband are $2.8 \pm 3.9$
events for the $X(1440)$, and $0.1\pm 1.5$ events for $f_{1}(1285)$,
which are neglected in the branching fraction measurements.

With the detection efficiencies determined from a phase space
distributed MC simulation of $\psi(3686)\rightarrow \omega X(1440)$
and $\psi(3686)\rightarrow \omega f_{1}(1285)$, which are
$(7.92\pm0.13)\%$ and $(8.02\pm0.13)\%$ respectively, the products of branching
fractions are determined to be

\begin{eqnarray}
  \mathcal{B}(\psi(3686) \rightarrow \omega X(1440))\cdot \mathcal{B}( X(1440) \rightarrow K^{+} K^{-} \pi^{0})
  & = &(1.09 \pm 0.20~({\rm stat.})) \times 10^{-5},\nonumber
\end{eqnarray}
\begin{eqnarray}
  \mathcal{B}(\psi(3686) \rightarrow \omega f_{1}(1285))\cdot \mathcal{B}( f_{1}(1285) \rightarrow K^{+} K^{-} \pi^{0})
  & = &(1.25 \pm 0.70~({\rm stat.})) \times 10^{-6}.\nonumber
\end{eqnarray}

Using the Bayesian approach, the upper limit on the branching fraction
for $\psi(3686)\rightarrow \omega f_{1}(1285)$ is determined.  The
upper limit on the number of $f_{1}(1285)$ events is $15$ at the
$90\%$ C.L..

\subsection{Branching fractions for $\psi(3686) \rightarrow \omega
  K^{*+} K^{-}+c.c.$}

\begin{figure}[htbp]
  \includegraphics[width=0.45\textwidth]{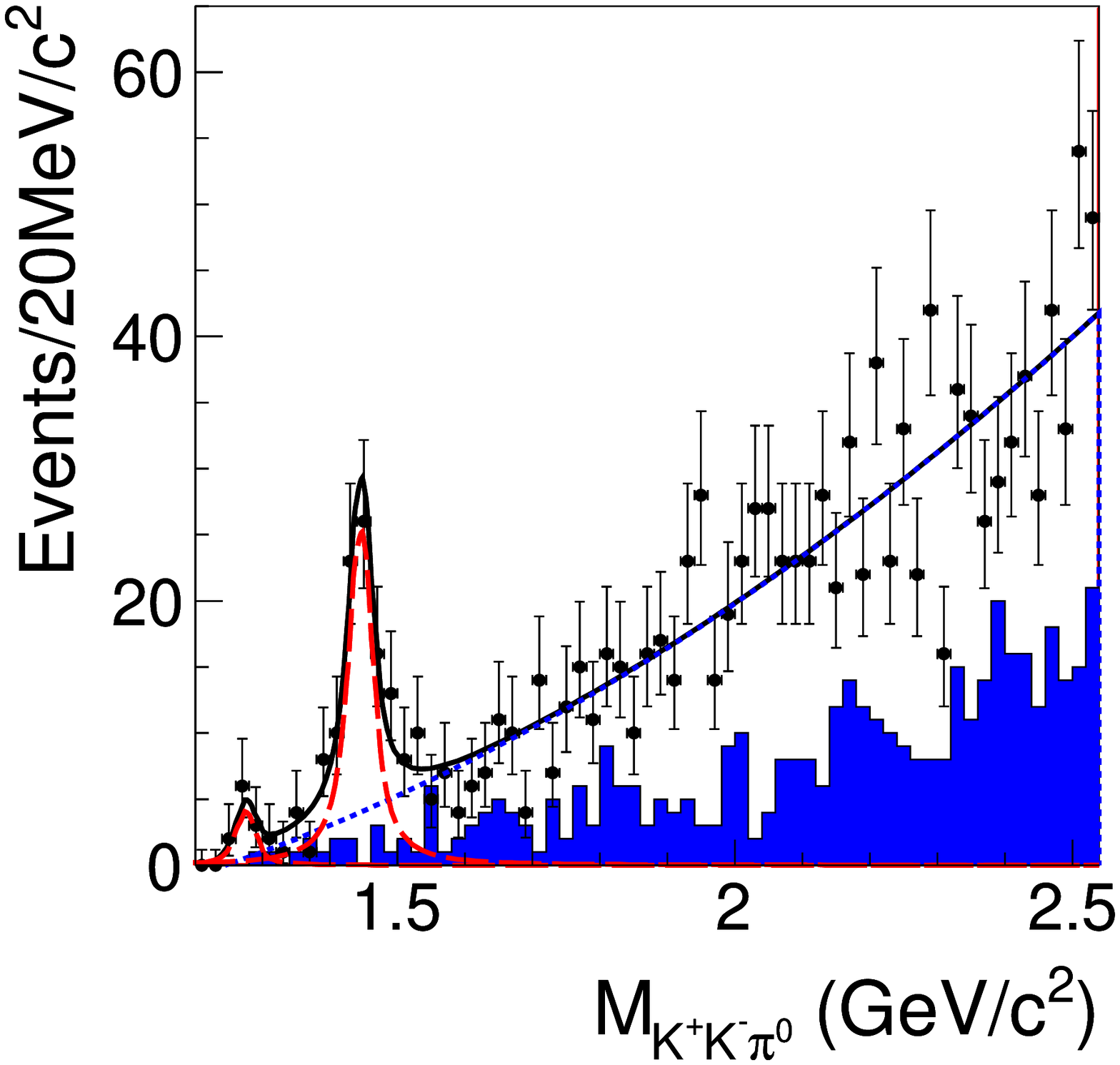}
  \includegraphics[width=0.45\textwidth]{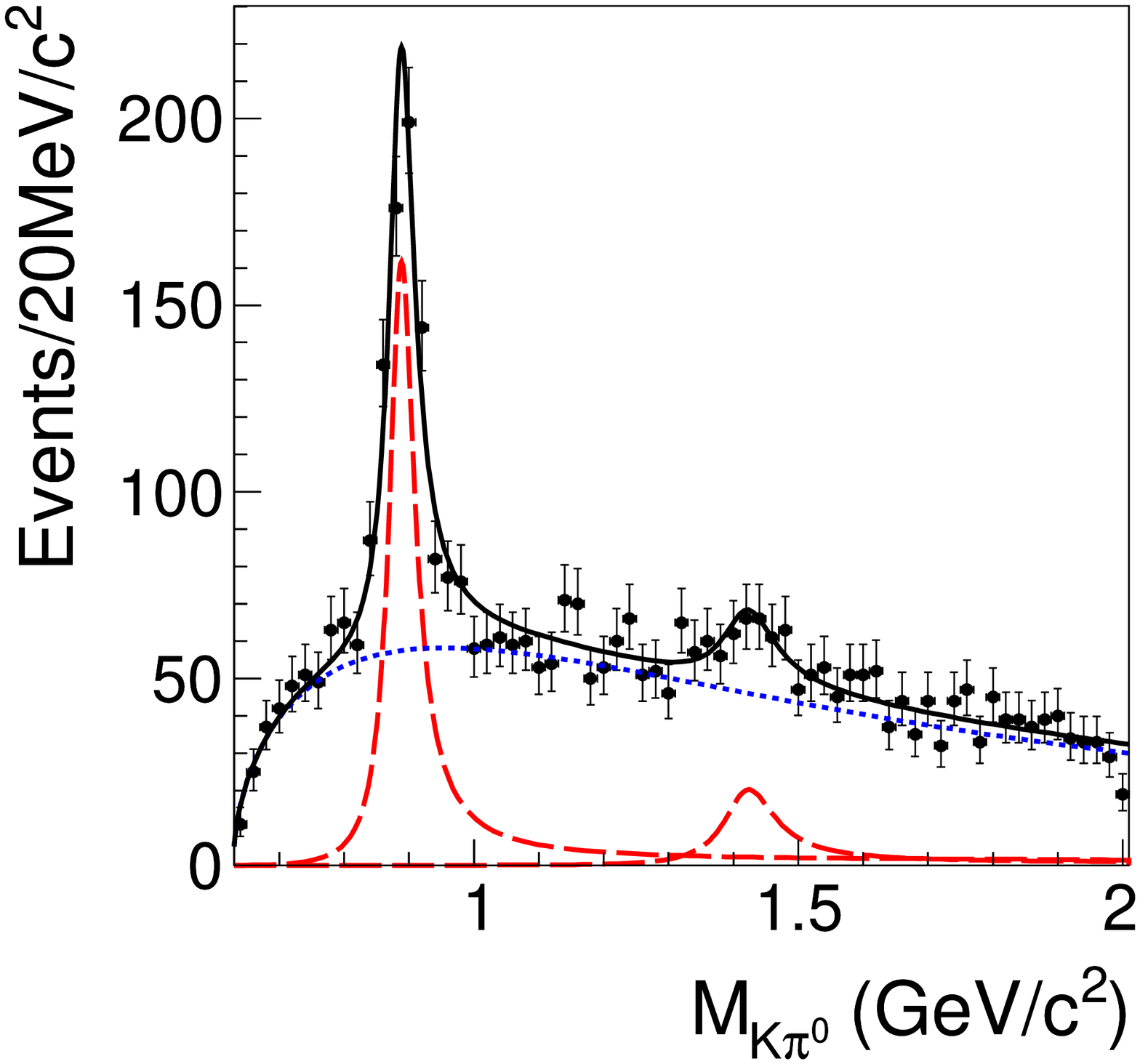}
  \put(-330,165){\bf~(a)} \put(-60,165){\bf~(b)}\\
  \caption{(a) The $M_{K^{+} K^{-}\pi^{0}}$ invariant-mass distribution
    for $\psi(3686) \rightarrow \omega K^{+}K^{-}\pi^{0}$ candidate events.
    The shaded histogram is for non-$\omega$ background events estimated from the
    $\omega$ sideband.
    (b) The combined mass spectra of $K^{+} \pi^{0}$ and $K^{-}
    \pi^{0}$ for $\psi(3686) \rightarrow \omega
    K^{*+}K^{-}$ candidate events.
    Solid curves are the fitting results; dashed lines are for
    the signal; and dotted lines are for the background.}
  \label{fig:x1440-wkkp0}
\end{figure}

The $M_{K^{\pm}\pi^{0}}$ distribution (both $M_{K^{+} \pi^{0}}$ and
$M_{K^{-} \pi^{0}}$ are included) is shown in
Fig.~\ref{fig:x1440-wkkp0}(b), where the $K^*(892)$ and
$K^{*}_{2}(1430)$ are clear. Using the same functions
as described in Section~\ref{sect-ppp}, an unbinned maximum likelihood
fit to $M_{K^{\pm}\pi^{0}}$ is performed. The fit yields
$678.8\pm65.3$ $K^{*\pm}(892)$ and $142.8\pm39.0$
$K^{*\pm}_{2}(1430)$ events with a statistical significance of
$4.5\sigma$. The mass and width of $K^{*\pm}(892)$ are determined to
be $M=889.6 \pm 2.1$ MeV$/c^{2}$ and $\Gamma=49.2\pm 5.5$ MeV/$c^2$.

The background contributions from non-$\omega$ processes estimated
by fitting $M_{K^{\pm}\pi^0}$ for events in the $\omega$
sideband are $(144.2 \pm 24.8)$ events for $K^{*\pm}(892)$,
which is subtracted from the total $K^{*\pm}(892)$ yield,
while no evident peak contributions for $K^{*\pm}_{2}(1430)$ are present.

The efficiencies determined from MC simulation are $(7.48\pm
0.07)\%$ and $(7.70\pm0.07)\%$ for $\psi(3686) \rightarrow \omega
K^{*+}(892) K^{-}$ and $\psi(3686) \rightarrow \omega
K^{*+}_{2}(1430) K^{-}$, respectively. The branching fractions are
determined to be
\begin{eqnarray}
  \mathcal{B}(\psi(3686) \rightarrow \omega K^{*+}(892) K^{-}) &=(2.26 \pm
  0.30~({\rm stat.})) \times 10^{-4},\nonumber
\end{eqnarray}
\begin{eqnarray}
  \mathcal{B}(\psi(3686) \rightarrow \omega K^{*+}_{2}(1430) K^{-}) &=(5.86
  \pm 1.61~({\rm stat.})) \times 10^{-5}.\nonumber
\end{eqnarray}

\section{Systematic uncertainties}~\label{systmatic-error-analysis}

Systematic errors in the branching fraction measurements mainly come
from the number of $\psi(3686)$ events, tracking, particle
identification, photon reconstruction, $K^{0}_{S}$ reconstruction,
kinematic fit, background estimation, signal shape and detection
efficiency.

The uncertainty on the number of $\psi(3686)$ is 0.81\% as reported
in Ref.~\cite{total-number-psi}. The tracking efficiencies have been
checked with clean channels such as $J/\psi \rightarrow \rho \pi$,
$J/\psi \rightarrow \pi^{+} \pi^{-} p \bar{p}$, $\psi(3686)
\rightarrow \pi^{+} \pi^{-} J/\psi$ ($J/\psi \rightarrow l^{+}
l^{-}$) and $J/\psi \rightarrow K^{*0}(892)K^{0}_{S}$
($K^{*0}\rightarrow  K^{\pm} \pi^{\mp}$, $K^{0}_{S} \rightarrow
\pi^{+} \pi^{-}$). It is found that the MC simulation agrees with
data within $1\%$ for each charged
track~\cite{photon-detector-dqgroup}. A $4\%$ systematic error due
to the MDC tracking efficiency is assigned for both $\psi(3686)
\rightarrow \omega K^{0}_{S} K^{+} \pi^{-}$ and $\psi(3686)
\rightarrow \omega K^{+} K^{-} \pi^0$. The uncertainty associated
with the kaon identification has been studied with a clean kaon
sample selected from $J/\psi\rightarrow K^{0}_{S}
K^{+}\pi^{-}$~\cite{photon-detector-dqgroup}. The data-MC efficiency
difference, which is $2\%$ per track, is assigned as the systematic
error for the kaon identification.

The uncertainty in the $K^{0}_{S}$ reconstruction
efficiency~\cite{bes3-paper}, which includes the geometric acceptance,
a pair of pion tracking efficencies and the $K^{0}_{S}$ secondary
vertex fit, is estimated with the decay of $J/\psi \rightarrow
K^{*0}K^{0}$, and the MC-data difference of $3.5\%$ is taken as the
systematic error for the $K^{0}_{S}$ reconstruction in the decay
$\psi(3686) \rightarrow \omega K^{0}_{S} K^{+} \pi^{-}$.  The photon
detection efficiency is studied with $J/\psi \rightarrow \rho^{0}
\pi^{0}$, and the difference between data and MC simulation is about
$1\%$~\cite{photon-detector-dqgroup} per photon. A $2\%$ ($4\%$)
systematic error for photon efficiency is used for the decay
$\psi(3686) \rightarrow \omega K^{0}_{S} K^{+}\pi^{-}$ ($\psi(3686)
\rightarrow \omega K^{+}K^{-}\pi^{0}$).

The uncertainties associated with the kinematic fit are estimated by
using $\psi(3686) \rightarrow \pi^{+} \pi^{-} J/\psi$ with $J/\psi
\rightarrow K^{0}_{S}K^{+}\pi^{-}\pi^{0}$ or $J/\psi \rightarrow
K^{+}K^{-}\pi^{0} \pi^{0}$ events. The efficiencies are obtained by
comparing the number of signal events with and without the kinematic
fit performed for data and MC simulation separately. A data-MC
difference of $5.4\%$ is found in $J/\psi \rightarrow
K^{0}_{S}K^{+}\pi^{-}\pi^{0}$ and $3.2\%$ in $J/\psi \rightarrow
K^{+}K^{-}\pi^{0} \pi^{0}$. The differences are taken as the
systematic errors.

The uncertainty in the $\mathcal{B}(\psi(3686) \rightarrow \omega
X(1440))$ measurement due to the background shape is estimated by
varying the background function from a 2nd-order polynomial to a
3rd-order polynomial in the fit to $M_{K\bar{K}\pi}$. The changes in
$X(1440)$ and $f_{1}(1285)$ yields, are $5.1\%$ ($3.2\%$) and
$2.5\%$ ($2.2\%$), respectively, in the decay $\psi(3686) \rightarrow \omega
K^{0}_{S}K^{+} \pi^{-}$ ($\psi(3686) \rightarrow \omega
K^{+}K^{-}\pi^{0}$). The uncertainty due to the fit
range is estimated by repeating the fits in the range [$1.2$,$2.4$]
GeV$/c^{2}$, and the differences of $6.1\%$ ($3.3\%$) and $3.3\%$
($6.0\%$) in the branching fractions for $X(1440)$ and $f_{1}(1285)$
in the decay $\psi(3686) \rightarrow \omega K^{0}_{S}K^{+}
\pi^{-}$ ($\psi(3686) \rightarrow \omega K^{+}K^{-}\pi^{0}$) are
assigned as systematic errors.

The uncertainty in the $\mathcal{B}(\psi(3686) \rightarrow \omega
X(1440))$ measurement due to the signal shape is considered to come
from the mass resolution, mass shift and phase space factor. By
varying the mass resolution by $\pm 0.5$ MeV/$c^2$ from the MC
expectation, the differences in $X(1440)$ yield, $1.0\%$ for
$K^{0}_{S}K^{+}\pi^{-}$ and $1.1\%$ for $K^{+}K^{-}\pi^{0}$,
are assigned as systematic errors respectively; the difference in
the $f_{1}(1285)$ yield can be ignored. By floating the mass of
$f_{1}(1285)$, the changes, $1.0\%$ for $K^{0}_{S} K^{+}
\pi^{-}$ and $2.1\%$ for $K^{+}K^{-}\pi^{0}$, are
taken as the systematic errors.
After taking the phase space factor into account, which depends
on the momentum of the $X(1440)$ in the $\psi(3686)$ rest frame and
the relative orbital angular momentum between the $\omega$ and
$X(1440)$, the differences in the fitting results of $X(1440)$ and
$f_{1}(1285)$  are found to be $2.7\%$ and $1.0\%$ in the
decay $\psi(3686) \rightarrow \omega K^{0}_{S} K^{+} \pi^{-}$, and
$0.5\%$ and $2.1\%$ in the decay $\psi(3686) \rightarrow \omega
K^{+} K^{-} \pi^{0}$, respectively, which are taken as the
uncertainties.

The selection efficiencies are determined from phase space distributed
MC simulations of the $\psi(3686)$ decay. The uncertainties in the
selection efficiencies are estimated by using efficiencies obtained
from MC samples that include intermediate states, or the angular
distribution associated with the $X(1440)$.  There is a small
difference ($2.5\%$ for $K^{0}_{S} K^{+}\pi^{-}$ and $3.2\%$ for
$K^{+} K^{-}\pi^{0}$) from the $\bar{K}^{*}K$ intermediate
process. The detection efficiencies are also checked by generating
$\psi(3686)\rightarrow \omega X(1440)$ MC events assuming X(1440) is a
pseudoscalar meson, and the differences for $X(1440)\rightarrow
K^{0}_{S}K^{+}\pi^{-}$ and $X(1440)\rightarrow K^{+}K^{-}\pi^{0}$ are
10.0\% and 11.3\%, respectively. To be conservative, the differences
of $10.0\%$ and $11.3\%$ are taken as the systematic errors for the
$X(1440) \rightarrow K^{0}_{S} K^{+}\pi^{-}$ and $X(1440) \rightarrow
K^{+}K^{-}\pi^{0}$, respectively.

The uncertainties in the $\mathcal{B}(\psi(3686) \rightarrow \omega
\bar{K}^{*}K)$ measurement due to the fit range are estimated to be
$7.6\%$ ($8.9\%$), $3.6\%$ ($9.1\%$) and $3.6\%$ ($11.6\%$) for the
decay $K^{*\pm}(892) (K^{*\pm}_{2}(1430))\rightarrow
K^{0}_{S}\pi^{\pm}$, $K^{*0}(892) (K^{*0}_{2}(1430))\rightarrow
K^{\mp}\pi^{\pm}$ and $K^{*\pm}(892) (K^{*\pm}_{2}(1430))\rightarrow
K^{\pm}\pi^{0}$, respectively. The differences by changing the sideband
range ($0.014$ GeV/$c^{2}<|M_{\pi^{+}\pi^{-}}- m_{K^{0}_{S}}|<0.022$
GeV/$c^{2}$ or $0.08$
GeV/$c^{2}<|M_{\pi^{+}\pi^{-}\pi^{0}}-m_{\omega}|<0.12$ GeV/$c^{2}$)
are estimated for the above decay modes. For $K^{*}(892)$,
the differences in the fitting results with or without the presence of
$K^{*}_{2}(1430)$ or by replacing $K^{*}_{2}(1430)$ with
$K^{*}_{0}(1430)$ (or $K^{*}(1410)$) are also estimated. The above
largest differences are taken as the systematic errors of
the $K^{*}(892)$.

All the contributions are summarized in
Table~\ref{table:systematic-error} and
Table~\ref{table:systematic-error-wkstark}. The total systematic uncertainty is given
by the quadratic sum of the individual errors, assuming all sources to be independent.

\begin{table}
\centering
  \caption{The systematic errors ($\%$) of $\mathcal{B}(\psi(3686) \rightarrow \omega X \rightarrow \omega K \bar{K} \pi)$.}
 \begin{tabular}{l|cc|cc}\hline
Sources & \multicolumn{2}{c|}{$\omega K^{0}_{S}K^{+}\pi^{-}$} & \multicolumn{2}{c}{$\omega K^{+}K^{-}\pi^{0}$}\\
  & $X(1440)$     & $f_{1}(1285)$   & $X(1440)$   & $f_{1}(1285)$ \\\hline
The number $\psi(3686)$ events & \multicolumn{2}{c|}{0.8} &  \multicolumn{2}{c}{0.8} \\
MDC tracking         &  \multicolumn{2}{c|}{4}      & \multicolumn{2}{c}{4} \\
Particle identification & \multicolumn{2}{c|}{2}     & \multicolumn{2}{c}{4} \\
$K^{0}_{S}$ reconstruction & \multicolumn{2}{c|}{3.5} & \multicolumn{2}{c}{-} \\
Photon efficiency & \multicolumn{2}{c|}{2}    & \multicolumn{2}{c}{4}\\
 Intermediate decays    & \multicolumn{2}{c|}{0.8} & \multicolumn{2}{c}{0.8} \\
Kinematic fit & \multicolumn{2}{c|}{5.4} & \multicolumn{2}{c}{3.2}\\
Background uncertainty      & 6.1 & 3.3 & 3.3 & 6.0     \\
Signal shape           & 2.9 & 1.4 & 1.2 & 3.0   \\
 MC Eff. Uncertainty    & 10.0 & - & 11.3 & - \\\hline
Total & 14.6 & 8.9 & 14.2  & 10.3  \\\hline
\end{tabular}
  \label{table:systematic-error}
\end{table}

\begin{table}
\centering
  \caption{The systematic errors ($\%$) of $\mathcal{B}(\psi(3686) \rightarrow \omega \bar{K}^{*} K)$.}
 \begin{tabular}{l|cccc|cc}\hline
 Sources                        & \multicolumn{4}{c|}{$\omega K^{0}_{S}K^{+}\pi^{-}$}
                        & \multicolumn{2}{c}{$\omega K^{+}K^{-}\pi^{0}$}\\
  & $K^{*\pm}(892)$  & $K^{*\pm}_{2}(1430)$ & $K^{*0}(892)$ & $K^{*0}_{2}(1430)$
  & $K^{*\pm}(892)$  & $K^{*\pm}_{2}(1430)$\\\hline
 The number of $\psi(3686)$ events &  \multicolumn{4}{c|}{0.8} & \multicolumn{2}{c}{0.8} \\
 MDC tracking         &  \multicolumn{4}{c|}{4}      & \multicolumn{2}{c}{4} \\
  Particle identification &\multicolumn{4}{c|}{2}     & \multicolumn{2}{c}{4} \\
  $K^{0}_{S}$ reconstruction & \multicolumn{4}{c|}{3.5} & \multicolumn{2}{c}{-}     \\
 Photon efficiency & \multicolumn{4}{c|}{2}    & \multicolumn{2}{c}{4}\\
 Intermediate decays    & \multicolumn{4}{c|}{0.8} & \multicolumn{2}{c}{0.8} \\
Kinematic fit & \multicolumn{4}{c|}{5.4} & \multicolumn{2}{c}{3.2}\\
 Background uncertainty      & 7.6 & 8.9 & 3.6 & 9.1 & 7.2 & 11.6    \\\hline
 Total & 11.2 & 12.1 & 9.0 & 12.3 & 10.6 &14.0 \\\hline
\end{tabular}

  \label{table:systematic-error-wkstark}
\end{table}

\section{\bf Discussion}

\begin{figure}[htbp]
  \includegraphics[width=0.45\textwidth]{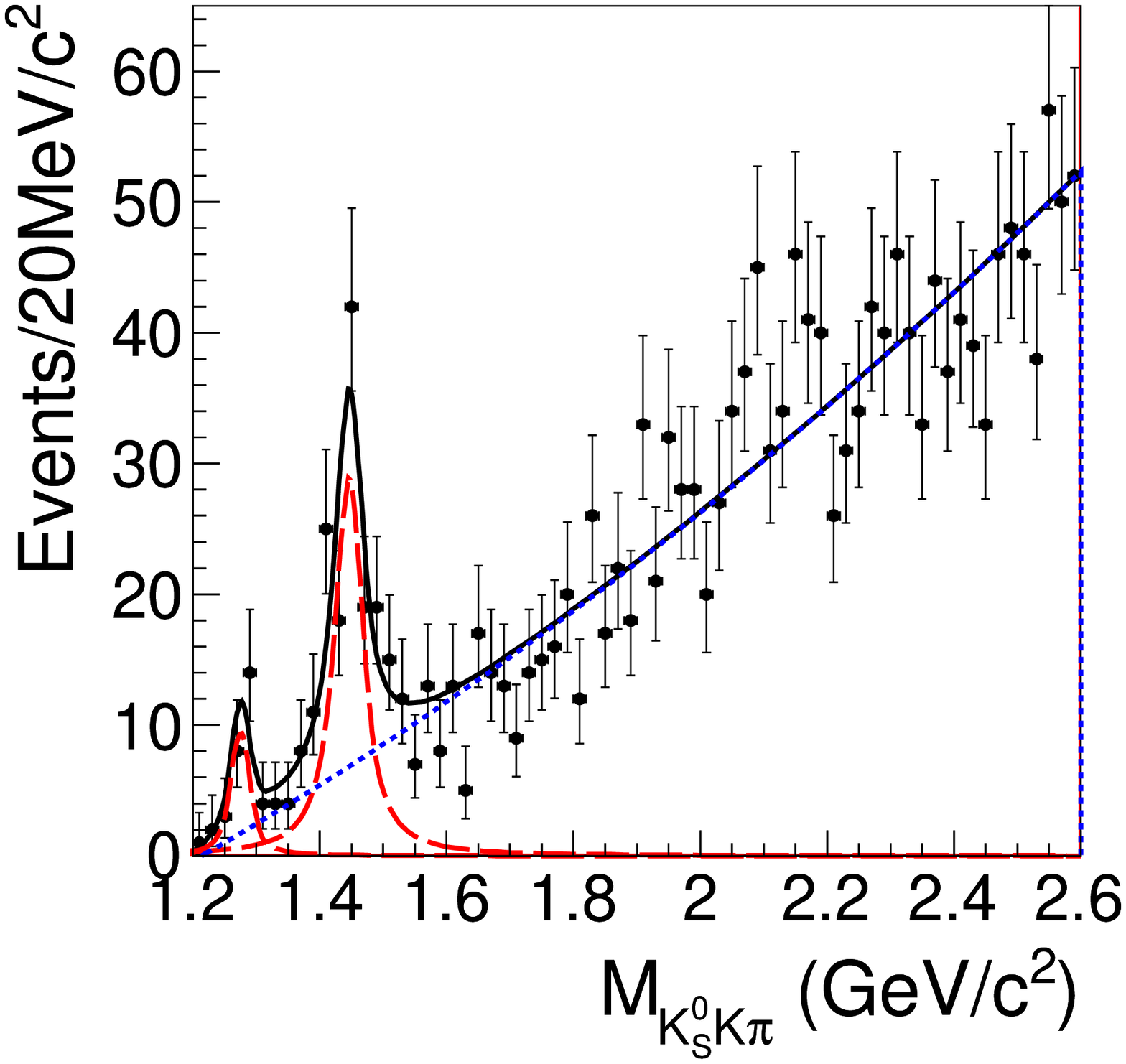}
  \includegraphics[width=0.45\textwidth]{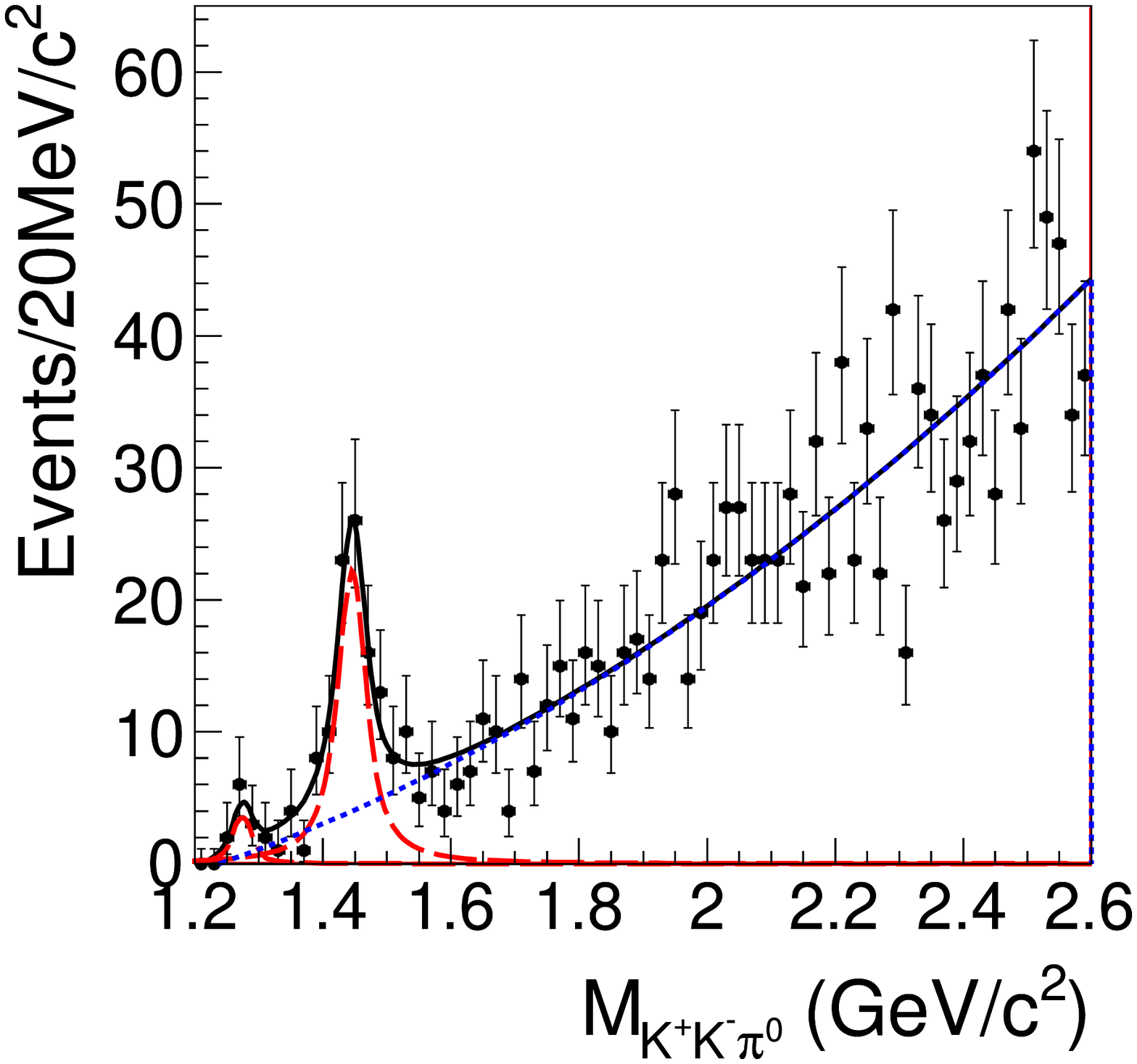}
  \put(-330,165){\bf~(a)} \put(-60,165){\bf~(b)}\\
\caption{ The simultaneous maximum-likelihood fit to the mass
spectra for $K^{0}_{S} K^{\pm}\pi^{\mp}$ (a) and $K^{+}K^{-}\pi^{0}$
(b). Points with error bars are data, and the curves show the best
fits.} \label{fig:simutaneous-x1440-kskp-kkp0}
\end{figure}

As the $X(1440)$ and $f_{1}(1285)$ are observed in both
$K^{0}_{S} K^{\pm}\pi^{\mp}$ and $K^{+}K^{-}\pi^{0}$ final states, a
simultaneous maximum-likelihood fit is performed to the mass spectra
to extract a more precise determination of the resonant parameters and
branching fractions. The fit includes three components, the
$X(1440)$, $f_1(1285)$, and background, as used in the fit to
each individual mode in Section~\ref{sect-x1440} and
Section~\ref{sect-x1440-kkp0}. The fit, shown in
Fig.~\ref{fig:simutaneous-x1440-kskp-kkp0}, has a
$\chi^{2}/ndf = 72.9/70 = 1.04$, and the statistical significances
of the $X(1440)$ and $f_{1}(1285)$ are $13.3\sigma$ and
$5.4\sigma$, respectively. The mass and width of the $X(1440)$ from
the fit are $M=1452.7 \pm 3.3$~MeV$/c^{2}$ and $\Gamma= 45.9 \pm
8.2$~MeV/$c^2$. The yields of the $X(1440)$ events are $111.4\pm 17.2$ in
the $K^{0}_{S} K^{\pm}\pi^{\mp}$ mode and $82.4 \pm 13.5$ in the
$K^{+}K^{-}\pi^{0}$ mode, while those of the $f_{1}(1285)$ events
are $23.1\pm 7.1$ in the $K^{0}_{S} K^{\pm}\pi^{\mp}$ mode and $8.7
\pm 4.6$ in the $K^{+}K^{-}\pi^{0}$ mode, in good
agreement with the separate fits to the two modes, as shown in
Table~\ref{table:sum-branching-mass1}. Combining the observed
numbers of signal events with efficiencies and taking properly into account the
Clebsch-Gordan coefficients, the branching
fractions are
\begin{eqnarray}
\mathcal{B}(\psi(3686) \rightarrow \omega X(1440))\cdot \mathcal{B}(
X(1440) \rightarrow K \bar{K} \pi) & = &(5.48 \pm 0.61~({\rm
stat.})\pm 0.86~({\rm sys.})) \times 10^{-5}, \nonumber
\end{eqnarray}
and
\begin{eqnarray}
\mathcal{B}(\psi(3686) \rightarrow \omega f_{1}(1285))\cdot
\mathcal{B}( f_{1}(1285) \rightarrow K \bar{K} \pi) & = &(8.78 \pm
2.33~({\rm stat.}) \pm 0.96~({\rm sys.})) \times 10^{-6}.\nonumber
\end{eqnarray}
The first errors are statistical and the second ones systematic. In
calculating the systematic errors, the correlations between the
errors in the two modes shown in Table~\ref{table:systematic-error}
are properly taken into account.

\begin{figure}[htbp]
  \centering
    \includegraphics[width=0.45\textwidth]{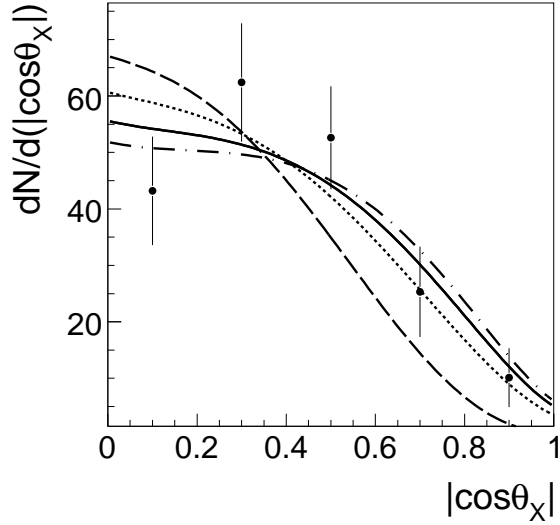}
\caption{The $|\cos\theta_{X}|$ distribution with the fit results;
solid line is for $1+\alpha \cos^{2}\theta_{X}$ with $\alpha=0.58\pm
0.64$; dash-dotted line is for $1+ \cos^{2}\theta_{X}$; dotted line is
for $1+ 0 \cdot \cos^{2}\theta_{X}$; and dashed line is for $1-
\cos^{2}\theta_{X}$. The expectation distributions are multiplied by
the efficiency.  }
  \label{fig:combine-kkp-angle}
\end{figure}

To examine the spin-parity of the events at around
$1.4$~GeV$/c^{2}$, we try to measure the polar angle distribution
(denoted as $\cos\theta_{X}$) of the $X(1440)$ in the $\psi(3686)$
rest frame. The $|\cos\theta_{X}|$ distribution is divided into five
bins in the region of $[0,1]$. In each bin, the combined
$K\bar{K}\pi$ mass spectrum of the two decay modes is fitted. For $X(1440)$,
the mass and width are fixed to those of the combined mass spectrum
fitting for the whole angular range; for $f_{1}(1285)$, the mass and
width are fixed at the known values. By repetition of the mass fit
in five bins of $|\cos\theta_{X}|$, the number of $X(1440)$ events can
be obtained. Figure~\ref{fig:combine-kkp-angle} shows the polar
angular distribution for signal yields, where the errors are
statistical only.

The angular distribution is fitted to $1+\alpha \cos^{2}\theta_{X}$,
as shown in the solid line in Fig.~\ref{fig:combine-kkp-angle}, and
$\alpha=0.58\pm 0.64$ is obtained with a probability of $29\%$.
Around 1.44~GeV/c$^2$, there are two known resonances, namely the
$\eta(1440)$ with $J^{PC}=0^{-+}$ and the $f_{1}(1420)$ with
$J^{PC}=1^{++}$~\cite{pdg-2012}. The present statistics are not
sufficient to establish a $1+\cos^{2}\theta_{X}$ behavior as a
pseudoscalar meson candidate of $\eta(1440)$, but the data clearly
favor $\alpha=1$ or $\alpha=0$ over $\alpha=-1$, as can be seen from
Fig.~\ref{fig:combine-kkp-angle}.

\section{\bf Summary}

With a sample of $106$ million $\psi(3686)$ events, the decay of
$\psi(3686) \rightarrow \omega K \bar{K} \pi$ is studied for the first
time. In addition to the mass enhancement ($X(1440)$) around
$1.44$~GeV/$c^{2}$, the $f_{1}(1285)$ is observed in the mass spectrum
of $K\bar{K}\pi$. From investigating the $X(1440)$ polar angle
distribution, the present statistics are not sufficient to establish a
$1+\cos^{2}\theta_{X}$ behavior as a pseudoscalar meson candidate of
$\eta(1440)$, but the data favor $\alpha=1$ or $\alpha=0$ over $\alpha=-1$.
The product branching fraction upper limit of $f_{1}(1285)$ in each
individual mode is also presented after taking into account the
systematic error by dividing by a factor $(1-\delta_{sys})$, where the
$\delta_{sys}$ is the systematic error for the corresponding
decay. Also the branching fractions of $\psi(3686) \rightarrow
\omega \bar{K}^{*}K$ for the charged and neutral mode are
measured for the first time. The observed $K^{*}(1430)$ favors
$K^{*}_{2}(1430)$ over $K^{*}_{0}(1430)$ and $K^{*}(1410)$. The
numbers of observed events, detection efficiencies and branching
fractions (or upper limits) are summarized in
Tables~\ref{table:sum-branching-mass1} and
\ref{table:sum-branching-mass2}.

To compare with the $12\%$ rule,
Tables~\ref{table:sum-branching-mass1}
and~\ref{table:sum-branching-mass2} also include the corresponding
$J/\psi$ branching fractions~\cite{omegakkbarpi2008}, as well as the
ratio $Q_{h}$. The data show that $\psi(3686)$ decaying into $\omega
X(1440)$ and $\omega \bar{K}^*(892)K$ are suppressed by a factor of
$2-4$.

\begin{table}
  \centering
\caption{The branching fractions and upper limits ($90\%$ C.L.) for
$\psi(3686)\rightarrow \omega X \rightarrow \omega K \bar{K} \pi$
decays. Results for corresponding $J/\psi$
decays~\cite{omegakkbarpi2008} and the ratio
$Q_{h}=\frac{\mathcal{B}(\psi(3686)\rightarrow
h)}{\mathcal{B}(J/\psi \rightarrow h)}$  are also given.}
  \begin{tabular}{l|ccccc}\hline
decay mode & $N_{sig}$ & $\epsilon(\%)$ & $\mathcal{B}(\psi(3686))
(\times 10^{-5})$ &  $\mathcal{B}(J/\psi) (\times 10^{-5})$ &
$Q_{h}(\%)$  \\\hline
$\omega X(1440) \rightarrow \omega K^{0}_{S} K^{+} \pi^{-}$ & $109 \pm 18$ & $10.41\pm0.14$ & $1.60 \pm 0.27 \pm 0.24$ & $48.6\pm 6.9 \pm 8.1$ & $3.3\pm1.1$\\
$~~~~~~~~~~~~~\rightarrow \omega K^{+} K^{-} \pi^{0}$ & $82 \pm 15$ & $7.92\pm0.13$ & $1.09 \pm 0.20 \pm 0.16$ &$19.2\pm5.7\pm3.8$ & $5.7\pm2.5$\\
$~~~~~~~~~~~~~\rightarrow \omega K \bar{K} \pi$       &   ...     &  ...           & $5.48 \pm 0.61\pm 0.86$   &  ...        & ...      \\
$\omega f_{1}(1285)\rightarrow \omega K^{0}_{S} K^{+} \pi^{-}$ & $21.6\pm7.0$ & $10.88\pm 0.15$ & $0.302 \pm 0.098 \pm 0.027$ & ... & ... \\
                        & $<31$ & $10.88\pm0.15$ & $<0.478$ & ... & ...\\
$~~~~~~~~~~~~~\rightarrow \omega K^{+} K^{-} \pi^{0}$ & $9.5\pm 5.3$ & $8.02\pm0.13$ & $0.125 \pm 0.070 \pm 0.013$ & ...&... \\
          & $<15$ & $8.02\pm0.13$ & $<0.221$ &... &... \\
$~~~~~~~~~~~~~\rightarrow \omega K \bar{K} \pi$ & ...  & ...  &
$0.878 \pm 0.233 \pm 0.096$ & ... &  ...      \\\hline
\end{tabular}
  \label{table:sum-branching-mass1}
\end{table}

\begin{table}
  \centering
\caption{The branching fractions of $\psi(3686) \rightarrow \omega
\bar{K}^{*}K$.}
  \begin{tabular}{l|cccccc}\hline
 decay mode  & $N_{sig}$ & $\epsilon(\%)$ & $\mathcal{B}(\psi(3686)) (\times 10^{-5})$ &  $\mathcal{B}(J/\psi) (\times 10^{-5})$
& $Q_{h}(\%)$  \\\hline
$\omega K^{*+}(892) K^{-} \rightarrow \omega K^{0}_{S}K^{+}\pi^{-}$       & $396.4\pm60.4$ & $9.58\pm0.08$ & $18.9\pm 2.9 \pm  2.2$  & $310\pm34\pm56$& $6.1\pm1.8$\\
$~~~~~~~~~~~~~~~~~~~\rightarrow \omega K^{+}K^{-}\pi^{0}$        & $534.6\pm69.9$ & $7.48\pm0.07$ &$22.6 \pm 3.0 \pm 2.4$ & $327\pm51\pm 68$& $7.0\pm2.2$\\
$\omega K^{*+}_{2}(1430) K^{-}\rightarrow \omega K^{0}_{S}K^{+}\pi^{-}$   & $128.5\pm30.0$  & $9.18\pm0.07$ & $6.39\pm 1.50 \pm  0.78$& ... & ... \\
$~~~~~~~~~~~~~~~~~~~~~\rightarrow \omega K^{+}K^{+}\pi^{0}$   & $142.8 \pm 39.0$ & $7.70\pm0.07$ & $5.86 \pm 1.61 \pm 0.83$ & ... & ...\\
$\omega \bar{K}^{*0}(892) K^{0} \rightarrow \omega K^{0}_{S}K^{+}\pi^{-}$   & $356.0\pm50.8$ & $9.66\pm0.08$ & $16.8\pm 2.5 \pm  1.6$  & $310\pm34\pm56$&$5.4\pm1.5$ \\
$\omega \bar{K}^{*0}_{2}(1430) K^{0} \rightarrow \omega
K^{0}_{S}K^{+}\pi^{-}$   & $115.7 \pm 41.3$ & $9.08\pm0.07$&
$5.82\pm 2.08 \pm  0.72$ &... &... \\\hline
\end{tabular}
  \label{table:sum-branching-mass2}
\end{table}

\section{\bf Acknowledgement}
The BESIII collaboration thanks the staff of BEPCII and the
computing center for their hard efforts. This work is supported in
part by the Ministry of Science and Technology of China under
Contract No. 2009CB825200; National Natural Science Foundation of
China (NSFC) under Contracts Nos. 10625524, 10821063, 10825524,
10835001, 10935007, 11125525, 11235011; Joint Funds of the National
Natural Science Foundation of China under Contracts Nos. 11079008,
11179007; the Chinese Academy of Sciences (CAS) Large-Scale
Scientific Facility Program; CAS under Contracts Nos. KJCX2-YW-N29,
KJCX2-YW-N45; 100 Talents Program of CAS; German Research Foundation
DFG under Contract No. Collaborative Research Center CRC-1044;
Istituto Nazionale di Fisica Nucleare, Italy; Ministry of
Development of Turkey under Contract No. DPT2006K-120470; U. S.
Department of Energy under Contracts Nos. DE-FG02-04ER41291,
DE-FG02-94ER40823; U.S. National Science Foundation; University of
Groningen (RuG) and the Helmholtzzentrum fuer Schwerionenforschung
GmbH (GSI), Darmstadt; WCU Program of National Research Foundation
of Korea under Contract No. R32-2008-000-10155-0.

\end{document}